\documentclass[journal, web, twocolumn]{IEEEtran}
\usepackage{cite}
\usepackage{amsmath,amssymb,amsfonts}
\usepackage{algorithmic}
\usepackage{graphicx}
\usepackage{makecell}
\usepackage{textcomp}
\usepackage{hyperref}
\usepackage{calligra}
\usepackage{multirow}
\usepackage{dblfloatfix}
\usepackage{flushend}

\hypersetup{
    colorlinks=true,
    linkcolor=magenta,
    filecolor=blue,      
    urlcolor=cyan,
    citecolor=green,
}

\def\BibTeX{{\rm B\kern-.05em{\sc i\kern-.025em b}\kern-.08em
    T\kern-.1667em\lower.7ex\hbox{E}\kern-.125emX}}
\begin{document}

\title{BUET Multi-disease Heart Sound Dataset: A Comprehensive Auscultation Dataset for Developing Computer-Aided Diagnostic Systems}
\author{Shams~Nafisa~Ali\textsuperscript{1}, Afia~Zahin\textsuperscript{1}, Samiul~Based~Shuvo\textsuperscript{1}, Nusrat~Binta~Nizam\textsuperscript{1, 2}, \\Shoyad~Ibn~Sabur~Khan~Nuhash\textsuperscript{1}, Sayeed~Sajjad~Razin\textsuperscript{1}, S.m.~Sakeef~Sani\textsuperscript{1}, Farihin~Rahman\textsuperscript{1}, \\
Nawshad~Binta~Nizam\textsuperscript{3}, Farhat Binte Azam\textsuperscript{1}, Rakib~Hossen\textsuperscript{1}, Sumaiya~Ohab\textsuperscript{1}, \\ Nawsabah~Noor\textsuperscript{4}~and~~Taufiq~Hasan\textsuperscript{1,5,†}
%
\thanks{\textsuperscript{1}These authors are with mHealth Lab, Department of Biomedical Engineering, BUET, Dhaka, Bangladesh.}
\thanks{\textsuperscript{2}Nusrat~Binta~Nizam is also affiliated with the Meinig School of Biomedical Engineering, Cornell University, Ithaca, NY.}
\thanks{\textsuperscript{3}Nawshad~Binta~Nizam is with Department of Bioengineering, University of Pittsburgh, PA.}
\thanks{\textsuperscript{4}Nawsabah~Noor is with the Popular Medical College, Dhaka, Bangladesh.}
\thanks{\textsuperscript{5, †}Taufiq Hasan is the corresponding author. He has a secondary affiliation with the Center for Bioengineering Innovation and Design, Department of Biomedical Engineering, Johns Hopkins University, Baltimore, MD. Email: taufiq@bme.buet.ac.bd, taufiq.hasan@jhu.edu.}
\thanks{This project was partially funded by the Kaggle Open Data Research Grant.}
}

\maketitle
\begin{abstract}
Cardiac auscultation, an integral tool in diagnosing cardiovascular diseases (CVDs), often relies on the subjective interpretation of clinicians, presenting a limitation in consistency and accuracy. Addressing this, we introduce the BUET Multi-disease Heart Sound (BMD-HS) dataset— a comprehensive and meticulously curated collection of heart sound recordings. This dataset, encompassing 864 recordings across five distinct classes of common heart sounds, is representative of a broad spectrum of valvular heart diseases, with a focus on diagnostically challenging cases. The standout feature of the BMD-HS Dataset is its innovative multi-label annotation system, which captures a diverse range of diseases and unique disease states. This system significantly enhances the dataset's utility for developing advanced machine learning models in automated heart sound classification and diagnosis. By bridging the gap between traditional auscultation practices and contemporary data-driven diagnostic methods, the BMD-HS Dataset is poised to revolutionize CVD diagnosis and management, providing an invaluable resource for the advancement of cardiac health research. The dataset is publicly available in this link:~\href{https://github.com/sani002/HS-Dataset}{https://github.com/mHealthBuet/BMD-HS-Dataset.}
\end{abstract}

 
\section{Introduction}
Cardiovascular diseases (CVDs) are at the forefront of global health concerns, consistently ranking as the leading cause of mortality worldwide~\cite{Di2024}. This group of disorders, encompassing coronary artery diseases, valvular heart diseases, and genetic cardiac conditions, presents a multifaceted challenge to healthcare systems. Alarmingly, forecasts predict an escalation in CVD prevalence, with an estimated 23 million deaths attributed to these diseases by 2030~\cite{bin2022prediction}. The severity of CVDs is further compounded by disparities in healthcare accessibility, particularly in developing and underdeveloped regions. The absence or scarcity of primary health facilities and specialized cardiac care, such as cardiology services, exacerbates the issue~\cite{oliveira2021circor}. This lack of infrastructure often leads to delayed diagnosis of heart conditions, adversely affecting community health outcomes.
\par 
In the realm of CVD diagnosis, heart sound analysis has emerged as a promising avenue for early detection~\cite{shuvo2021cardioxnet}. Despite its potential, the reliability of this method is hindered by subjective interpretations in auscultation, a challenge more pronounced in resource-limited settings~\cite{alam2010cardiac}. The diagnostic accuracy of even trained cardiologists in such environments is notably limited. In response to these diagnostic challenges, there has been a growing interest in the application of artificial intelligence (AI), specifically machine learning and deep learning techniques, in the analysis of phonocardiogram (PCG) signals~\cite{ren2022deep, ma2023parameter, ali2023end}. These advanced computational methods have shown promise in enhancing the accuracy of CVD detection, both in one-dimensional time-domain PCG signals~\cite{shuvo2021cardioxnet, ma2023parameter, zhang2023multi} and in two-dimensional time-frequency representations~\cite{azam2021heart, qiao2022hs}. Nevertheless, the development of a robust AI-based diagnostic tool is dependent on the availability of comprehensive and well-annotated data.

The current landscape of available PCG datasets includes the PhysioNet/CinC Challenge 2016 Dataset~\cite{data_physionet}, Michigan Heart Sound and Murmur Database~\cite{michigan_data}, Github Open-access Dataset~\cite{data_git}, Heart Sounds Shenzhen (HSS) Dataset~\cite{data_HSS}, CirCor DigiScope Dataset~\cite{oliveira2021circor} and the PASCAL Heart Sound Challenge 2011 Dataset~\cite{data_pascal}. Although these datasets vary in their scope and depth, many of them lack comprehensive information on heart sound evaluations, including the identification of murmurs, extra sounds, severity, and disease categorizations~\cite{data_physionet, data_pascal}. The datasets also differ in the quality of recordings. While some of them are limited in number~\cite{michigan_data} and excessively noisy to be established as ground truths~\cite{data_pascal}, some are extremely clean and idealistic~\cite{data_git}. Both of these cases inadequately reflect real-world conditions, thereby being not very useful for the learning of diagnosis. Furthermore, there is a notable scarcity of data where multiple disease classes co-exist. For instance, in many patients, aortic stenosis (AS) and aortic regurgitation (AR) are simultaneously found although they may vary in severity~\cite{unger2020aortic}. When such cases occur, the algorithms may end up categorizing them under one particular class, that too with less confidence score. Again, none of the datasets present gold-standard verification of the disease categories via echocardiogram. In this work, we have devised a structured framework to produce a novel clinically-sourced dataset. Key contributions of this dataset include:
\par 

\begin{enumerate}
    \item \textbf{Multi-Label Annotations for Enhanced Learning:} The dataset incorporates multi-label annotations to capture both multiple diseases and unique disease states. This feature significantly enhances the dataset's value for complex supervised learning tasks, allowing for more nuanced and accurate disease classification.

    \item \textbf{Standardized Data Collection for Consistency and Bias Reduction:} All recordings are of uniform duration and are collected using the same stethoscope. This eliminates the scope of device or domain biases.  Furthermore, the homogenization of recording positions/sites ensures consistent data acquisition, crucial for maintaining data integrity and reliability.

    \item \textbf{Echocardiogram-Confirmed Diagnoses:} The dataset's credibility is strengthened by the validation of patient conditions via echocardiograms, making it particularly reliable for studying specific cardiovascular disorders. Additionally, the possibility of extracting textual data from echocardiogram reports using Natural Language Processing (NLP) adds a layer of depth to the dataset, enhancing its contextual understanding and clinical relevance.

    \item \textbf{Comprehensive Metadata Integration:} The dataset incorporates detailed metadata, such as the patient's profession, age, gender, smoking habits, and place of residence. This inclusion allows for the exploration of potential correlations between various types of CVDs and corresponding demographic and lifestyle factors, offering a more holistic understanding of CVD risk profiles.
\end{enumerate}
 
The paper is structured as follows: Section II overviews prevalent heart sounds and cardiovascular diseases. Section III details existing public heart sound datasets. In Section IV, methodologies for data collection, including subject selection, ethical considerations, participant demographics, instrumentation, and label annotation approaches, are discussed. Section V covers the design and characteristics of the new BMD-HS dataset. Sections VI and VII focus on dataset analysis and evaluation, including signal quality assessment and data distribution visualization. Concluding the paper, Sections VIII, IX, X, and XI address challenges, dataset impact and usability, benchmarking, and primary study conclusions, respectively.

\begin{figure}[!h]
    \centering
    \includegraphics[width=0.45\textwidth]{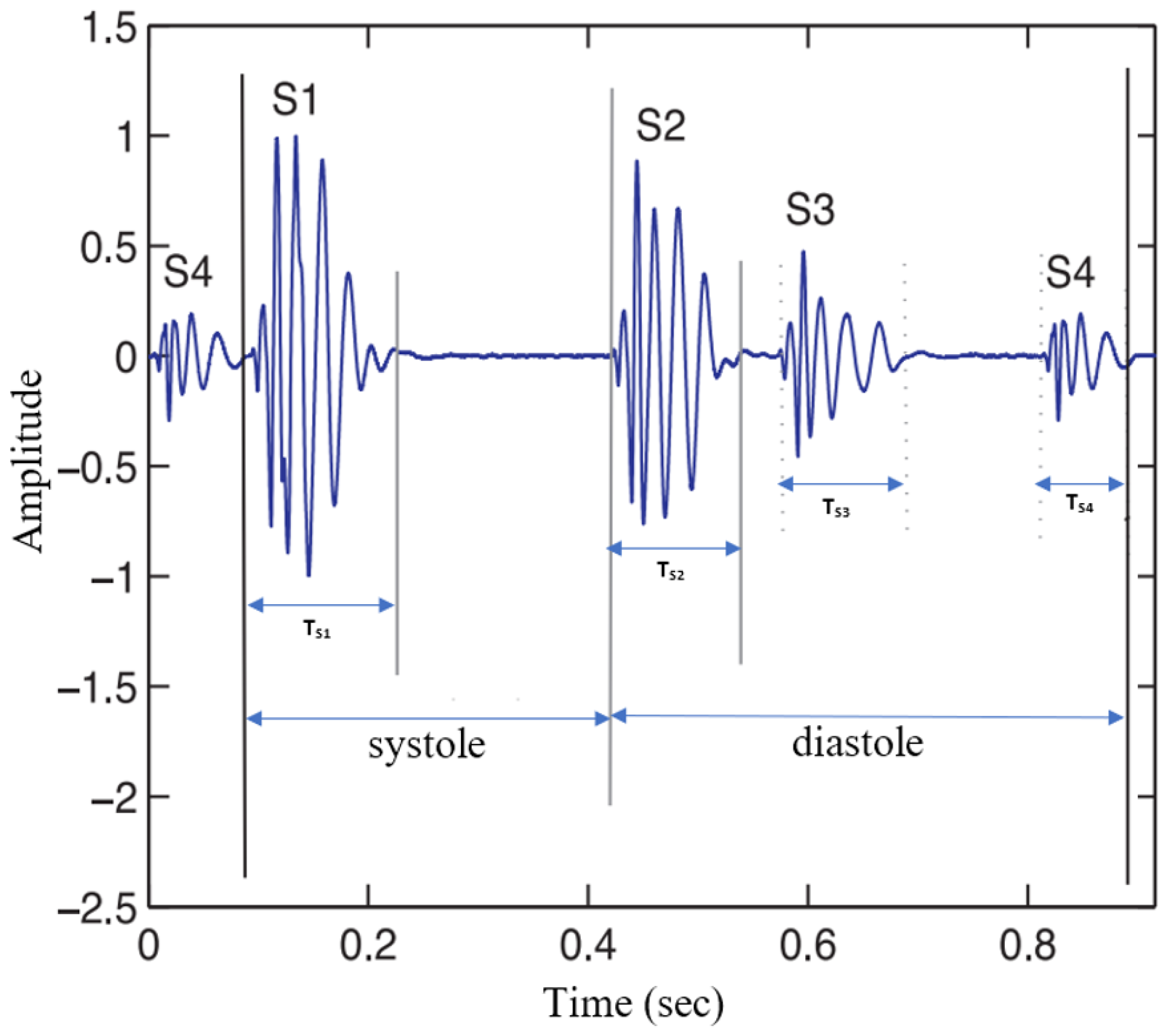}
    \caption{A Typical PCG Signal Displaying Heart Sounds (S1, S2, S3, S4) with Systole and Diastole Phases~\cite{Harimi_2022}}
    \label{signal}
\end{figure}

\section{Background}
\subsection{Cardiac Auscultation}
A vital part of the physical examination is cardiac auscultation, which uses a stethoscope to listen to the heart sounds. It is an essential clinical skill that allows medical professionals to evaluate a patient's heart health.

During auscultation, medical professionals listen for the heart's distinctive noises, often known as the "lub" and "dub," which indicate that the heart valves are closing. These sounds are termed S1 and S2. In addition to these normal sounds, clinicians may detect additional heart sounds, such as S3 and S4, or murmurs, which can indicate various cardiac conditions. These sounds are explained in detail in Table~\ref{table:heart_sounds}.

\begin{table}[!h]
\centering
\caption{Description of Heart Sounds and Murmurs}
\label{table:heart_sounds}
\renewcommand{\arraystretch}{1.5} 
\begin{tabular}{>{\centering\arraybackslash\bfseries}m{3cm}|>{\centering\arraybackslash}m{4cm}}
\hline
Heart Sound & \textbf{Description} \\
\hline
S1 (First heart sound) & Associated with the closure of the mitral and tricuspid valves at the beginning of systole. \\
\hline
S2 (Second heart sound) & Related to the closure of the aortic and pulmonary valves at the onset of diastole. \\
\hline
S3 (Third heart sound) & Can suggest heart failure or volume overload when present in adults, but may be a normal finding in children and young adults. \\
\hline
S4 (Fourth heart sound) & Usually indicative of a stiff or hypertrophic ventricle. \\
\hline
Murmurs & Produced by turbulent blood flow, which can be due to stenotic or regurgitant valves, septal defects, or other pathologies. \\
\hline
\end{tabular}
\end{table}




In Fig.~\ref{signal} a typical PCG signal is shown with heart sounds and the systolic and the diastolic periods are also displayed and identified.

\begin{figure}
    \centering
    \includegraphics[width=1\columnwidth]{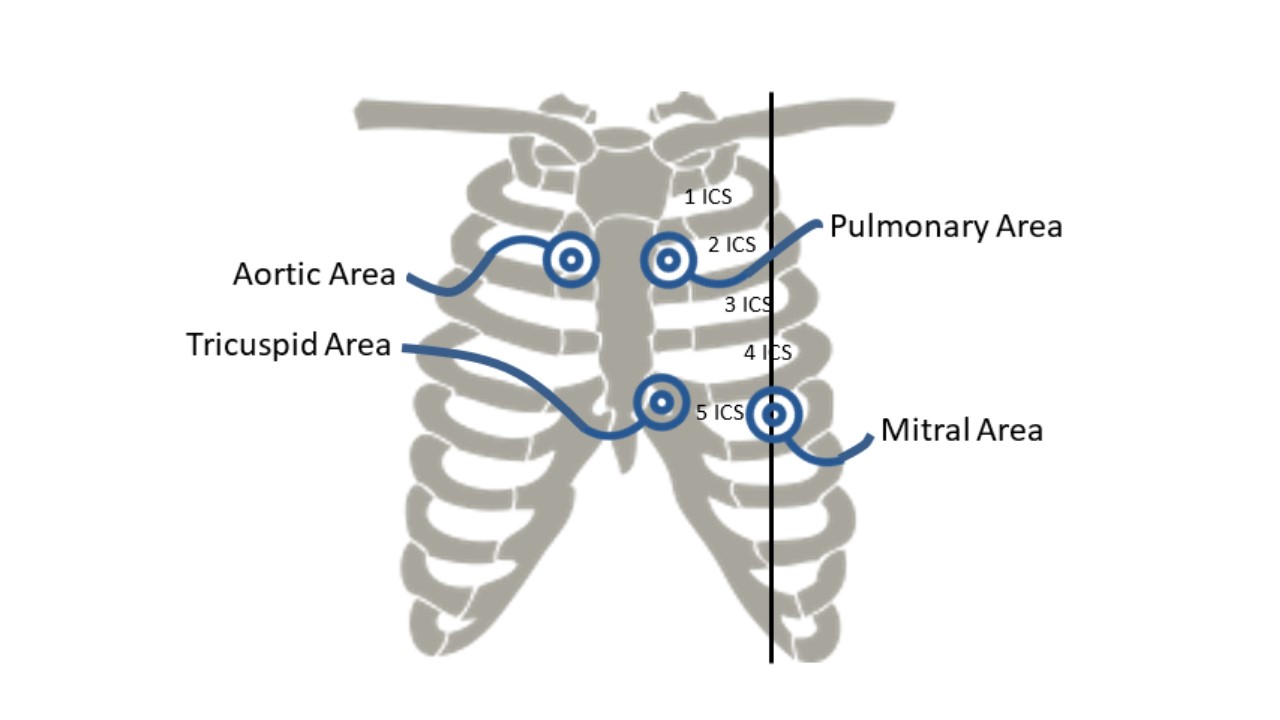}
    \caption{Heart Auscultation Collection Positions\cite{handee2020lexical}}
    \label{fig:uscultation}
\end{figure}

Auscultation is systematically performed in specific areas of the chest corresponding to the anatomical locations of the heart valves, as explained in Table~\ref{table:valves}. For optimal results, the stethoscope must be accurately positioned at these designated points on the patient's chest, as depicted in Fig.~\ref{fig:uscultation}.

\begin{table}[h!]
\centering
\caption{Auscultation Areas for Heart Valves}
\label{table:valves}
\renewcommand{\arraystretch}{1.5} 
\begin{tabular}{>{\centering\arraybackslash\bfseries}m{3cm}|>{\centering\arraybackslash}m{4cm}}
\hline
Valve & \textbf{Auscultation Area} \\
\hline
Aortic Valve & Right second intercostal space at the right sternal border. \\
\hline
Pulmonic Valve & Left second intercostal space at the left sternal border. \\
\hline
Tricuspid Valve & Lower left sternal border at the fourth or fifth intercostal space. \\
\hline
Mitral Valve & Fifth intercostal space at the midclavicular line. \\
\hline
\end{tabular}
\end{table}


\par
The first heart sound (S1), commonly known as "lub," is a crucial cardiovascular acoustic event that is essential to the clinical evaluation of cardiac function. S1 is produced as a consequence of the mitral and tricuspid valves, also known as M1 and T1, closing sequentially. Due to the quick succession of events, even while T1 occurs slightly later than M1, this temporal difference is usually imperceptible by auscultation, leading to the perception of the two components as a single sound.\cite{rijnberk1995medical}
The anatomical features of the atrioventricular (AV) valves are inextricably linked to the auditory properties of S1. The bigger size of the AV valves is the reason for the lower frequency of the S1 sound compared to the S2 heart sound. Unlike the diaphragm, which is better suited for higher-pitched noises, the low pitch of S1 makes it easier to detect with the stethoscope's bell, which is more sensitive to similar frequencies.

Furthermore, the anatomical intricacy of the AV valves results in a longer vibration, which lengthens the duration of S1 compared to S2, which is caused by the closure of the semi-lunar valves. Because abnormalities in the timing, pitch, or quality of the sound may signal pathological alterations in heart function, the clinical value of S1 features resides in their potential as a diagnostic tool.

\subsection{Prevalent Forms of Cardiovascular Diseases}
Valvular heart disease is an increasingly significant contributor to global cardiovascular morbidity and mortality, showing diverse and changing geographic patterns. Aortic valve stenotic disease is the predominant valvular pathology in developed countries, impacting approximately 9 million individuals globally~\cite{aluru2022valvular}. Its prevalence is on the rise due to an aging population and an increased incidence of atherosclerosis. Aortic valve diseases contribute to 61\% of all valvular heart disease-related deaths, while mitral valve diseases account for 15\%~\cite{mensah2019global}. The occurrence of aortic regurgitation, linked to diastolic hypertension, has also increased in developed nations. Mitral regurgitation affects around 24 million people worldwide, exhibiting considerable variation across different regions and countries.

Like the other South-East Asian developing countries, Bangladesh currently lacks a population-based surveillance system to monitor chronic non-communicable diseases due to inadequate health infrastructure development. There have only been a few studies done on the prevalence of CVD in Bangladeshi people\cite{van2008self}. To build a comprehensive dataset based on the most common heart diseases following the current demographics of Bangladesh, we emphasized four diseases which are: Aortic Stenosis (AS), Aortic Regurgitation (AR), Mitral Stenosis (MS), Mitral Regurgitation (MR), and their co-existence. These valvular diseases are mostly related to cardiac murmurs. Heart murmurs are often benign. These types of murmurs are called innocent murmurs. Still, there are some outliers. One possible cause of murmurs is an overworked or damaged heart valve. Some individuals have valve issues from birth. Others get them as a part of aging or from other heart problems~\cite{goldman2012approach}. The characteristic description of murmur with respect to each of the common valvular diseases considered is explained:

\begin{enumerate}
\item \textbf{Aortic Stenosis (AS):} Aortic stenosis is typically characterized by a high-pitched, "diamond-shaped" crescendo-decrescendo, mid-systolic ejection murmur that radiates to the neck and carotid arteries, and is best heard at the right upper sternal border~\cite{AorticSt72:online}. Early systole is when the murmur peaks in moderate aortic stenosis. But as the illness worsens, the peak occurs later in systole because it takes longer for the heart to empty and the aortic valve to close. 
\item \textbf{Aortic Regurgitation (AR):} The physical examination is essential for determining the origin of aortic regurgitation as well as any potential consequences, such as heart failure. The most common murmur in a patient with aortic regurgitation is a decrescendo early-diastolic blowing murmur, which is best heard at the third and fourth intercostal spaces on the left lower sternal border. Aortic root dilatation may be the reason for aortic regurgitation if this murmur is audible more on the right sternal border. This usually happens when there is an aortic dissection or ascending aortic aneurysm~\cite{AorticRe0:online}.
\item \textbf{Mitral Stenosis (MS):} During a physical examination, the length of the first part of the murmur and the location of the opening snap in diastole can be used to determine the degree of mitral stenosis~\cite{MitralSt83:online}. The term "rumble" refers to the low-frequency murmur caused by mitral stenosis. The pressure gradient between the left atrium and the left ventricle is reflected in the initial portion of the mitral stenosis murmur. It starts with the opening snap after S2 progresses to a decrescendo, and ends in mid-diastole. 
\begin{table*}[!hb]
\caption{A SUMMARY OF THE CURRENTLY AVAILABLE OPEN ACCESS DATASETS}
\label{datasets}
\centering
\renewcommand{\arraystretch}{1.5} 
\begin{tabular}{>{\centering\arraybackslash}p{2.3cm}|>{\centering\arraybackslash}m{0.7cm}|>{\centering\arraybackslash}m{3cm}|>{\centering\arraybackslash}m{1.4cm}|>{\centering\arraybackslash}m{1.2cm}|>{\centering\arraybackslash}m{1.2cm}|>{\centering\arraybackslash}m{1.5cm}|>{\centering\arraybackslash}m{2.8cm}}
\hline
\textbf{Dataset} & \textbf{Year} & \textbf{Classes} & \textbf{Total Recordings} & \textbf{Duration (s)} & \textbf{Age (Years)} & \textbf{Sampling Frequency (Hz)} & \textbf{Data Collecting Device} \\ \hline
\textbf{PhysioNet/CinC Challenge Dataset~\cite{liu2016open}} & 2016 & Normal, abnormal, unsure & 2435 & - & - & - & Collected from different sources \\ \hline       
\textbf{Circor DigiScope Dataset~\cite{oliveira2021circor}} & 2021 & Timing, shape, pitch, grading, quality, and location of each murmur & 5282 & 5-168 & - & 4000 & Littmann 3200 stethoscope \\ \hline
\textbf{Pascal Challenge Dataset~\cite{gomes2013classifying}} & 2011 & Normal, murmur, extrasystole, artifact & 656 & 1-30 & 1-17 & 4000 & Digital stethoscope \\ \hline
\textbf{HSCT-11~\cite{spadaccini2013performance}} & 2016 & Normal, abnormal & 412 & 6-36 & 20-70 & 11025 & ThinkLabs Rhythm digital electronic stethoscope \\  \hline
\textbf{Yaseen Khan's Dataset~\cite{yaseen2018classification}} & 2018 & Normal, aortic stenosis, mitral valve prolapse, mitral stenosis, mitral regurgitation & 1000 & 6-36 & Nearly 3 & 8000 & Collected from different sources \\  \hline
\textbf{EPHNOGRAM~\cite{kazemnejad2021ephnogram}} & 2021 & Normal, abnormal & 69 & 23-29 & 30-1800 & 8000 & Indoor fitness centre equipment \\ \hline
\textbf{Heartwave~\cite{alrabie2023heartwave}} & 2023 & Normal, abnormal with 8 distinct heart diseases & 1353 & 23-29 & - & - & Composed digital stethoscope and a data collection app \\  \hline
\end{tabular}
\end{table*}
\item \textbf{Mitral Regurgitation (MR):} When the mitral valve is incompetent, the left ventricle (LV) flows into the left atrium during ventricular systole, a condition known as mitral regurgitation~\cite{HeartMur73:online}. It is said that the high-pitched, "blowing" holosystolic murmur associated with mitral regurgitation is audible only near the apex. The murmur typically radiates to the axilla, though the exact direction of radiation varies on the type of mitral valve disease. To differentiate MR from tricuspid regurgitation, note that the MR murmur's intensity does not rise with inspiration.
\end{enumerate}

\section{Available Heart Sound Datasets}
In our comprehensive analysis, six heart sound datasets were meticulously selected based on a set of stringent criteria: complete data availability, online accessibility, relevance to the field, and comprehensive information on study populations and audio recording specifics. These datasets are succinctly summarized in Table~\ref{datasets}, which also juxtaposes key features from our proposed dataset for a comparative perspective.

A notable aspect of these existing datasets is the diversity in sampling frequencies, a parameter significantly influenced by the recording equipment used. Common frequencies range from 4 kHz to as high as 48 kHz, encompassing intermediate frequencies like 8 kHz, 11.025 kHz, 22.05 kHz, and 44.1 kHz. Although advanced sound cards capable of high-quality recording are utilized, it's crucial to note that oversampling beyond the Nyquist rate does not contribute additional pertinent information for signal processing or machine learning applications. Instead, it merely escalates the computational burden.

Consequently, a sampling frequency ranging from 2 kHz to 4 kHz is adequately sufficient for both human and automated machine-based diagnosis of heart sounds. This is contingent upon the implementation of an effective anti-aliasing analog filter to mitigate any undesirable spectral effects. This approach balances the need for quality and efficiency in heart sound analysis, streamlining the process for both clinical and research applications.

\subsection{2016 PhysioNet/CinC challenge Dataset\cite{liu2016open}}
The dataset provided for the 2016 PhysioNet/Computing in Cardiology (CinC) Challenge was designed for the task of classifying heart sounds into normal and abnormal categories. This database combined nine independent databases. The data, which includes 2435 heart sound recordings from 1297 patients overall, has been split into training and testing sets. The PCG signals last anywhere from 8 to 312.5 seconds. given that various devices and sampling rates were used to collect the data. The sounds were classified as normal, abnormal, and unsure after being recorded in both clinical and non-clinical settings. For every record, except for the unsure class, annotations indicating the locations of the basic heart sounds were supplied.

\subsection{Circor DigiScope Dataset\cite{oliveira2021circor}}

Oliveira et al. presented a pediatric dataset containing 29 heart sound recordings obtained from 29 patients, with ages ranging from six months to 17 years. The recordings vary in duration, with the shortest being around 2 seconds, the average being approximately 8 seconds, and the longest lasting about 20 seconds. These heart sounds were captured at Real Hospital Português in Recife, Brazil, using a Littmann 3200 stethoscope equipped with DigiScope Collector technology. The recordings were made at a sampling rate of 4 kHz, specifically from the mitral point. The dataset's uniqueness is primarily due to its extensive annotation which include murmur timing (early systolic/diastolic, mid systolic/diastolic, late systolic/diastolic, holosystolic), quality (musical, blowing, harsh), grading (I-VI), shape (crescendo, decrescendo, diamond, plateau) and pitch (high, medium, low).

\subsection{Pascal Challenge Dataset\cite{gomes2013classifying}}

This database consists of two sets of data. Dataset A was obtained from a population of undisclosed size using a smartphone application. Dataset B was collected at the Maternal and Fetal Cardiology Unit at Real Hospital Português (RHP) in Recife, Brazil, utilizing a digital stethoscope system. It encompasses a total of 656 recordings of heart sounds from an unspecified number of patients. The duration of the PCG signals ranges from 1 to 30 seconds, and they were recorded at a sampling rate of 4000 Hz. In dataset A, the sounds were captured from the apex point of volunteer subjects, while in dataset B, they were obtained from four cardiac auscultation locations on both healthy and unhealthy children.

\subsection{HSCT-11 (2016)\cite{spadaccini2013performance}}
This dataset is designed for evaluating the performance of biometric systems based on auscultation. It includes recordings from 206 patients, resulting in a total of 412 heart sounds. These sounds were collected from four auscultation locations: mitral, pulmonary, aortic, and tricuspid. The data was captured using the ThinkLabs Rhythm digital electronic stethoscope, which operates at a sampling rate of 11,025 Hz and provides a resolution of 16 bits per sample. However, it's worth noting that no specific information is provided regarding the health condition of each subject in this dataset.

\subsection{Yaseen Khan's Dataset\cite{yaseen2018classification}}
Yaseen Khan's dataset has 1000 audio files in .wav format. There are five classes of PCG recordings from various sources among which one is normal and four are abnormal classes. Each class has 200 audio files. The five classes are AS, MR, MS, Mitral Valve Prolapse (MVP), and N.

\subsection{EPHNOGRAM: A Simultaneous Electrocardiogram and Phonocardiogram Database\cite{kazemnejad2021ephnogram}}
The "EPHNOGRAM" database is a collection of simultaneous recordings of both Electrocardiogram (ECG) and PCG signals. It focused on creating cost-effective and energy-efficient devices capable of simultaneously recording ECG and PCG data on a sample-by-sample basis. The resulting database consists of 69 records obtained from 24 healthy young adults, aged between 23 and 29 (with an average age of 25.4 ± 1.9). These recordings were made during 30-minute stress-test sessions encompassing resting, walking, running, and biking activities, utilizing indoor fitness center equipment. In some records, additional auxiliary audio channels were used to capture environmental noises. The dataset is particularly valuable for conducting simultaneous multi-modal analyses of ECG and PCG signals. While no manual segmentation has been performed, the database provides a MATLAB code that can automatically and accurately detect all the R-peaks in the ECG signals. These R-peaks serve as reference points for locating the S1 and S2 components of the PCG.

\subsection{HeartWave~\cite{alrabie2023heartwave}}
The HeartWave dataset is a comprehensive collection of heart sound recordings that encompasses nine distinct classes of the most common heart sounds found in various classes and subclasses of cardiovascular diseases. In total, the dataset comprises 1353 recordings of heart sounds. In terms of patient distribution, the heart sound recordings came from 401 healthy people and 952 from various diseased people. It was developed by King Abdul-Aziz University and three other hospitals that offer specialized cardiovascular healthcare to the National Heart Institute in Cairo, Egypt.

\section{Study Design and Data Acquisition}
The development of dedicated algorithms for the preliminary screening of prevalent cardiovascular diseases (CVDs), particularly valvular diseases in Bangladesh, motivated the creation of the BMD-HS Dataset. A common challenge in medical datasets is the inherent bias toward a large representation of normal, healthy control subjects, resulting in a pronounced class imbalance among diseased categories. To address this issue and create a balanced dataset in terms of both number and gender, data collection targeted 100 subjects, including 20 healthy controls and 20 recordings from each of four distinct categories of commonly observed valvular disease classes:
\begin{itemize}
  \item Aortic Stenosis 
  \item Aortic Regurgitation 
  \item Mitral Regurgitation 
  \item Mitral Stenosis 
\end{itemize}

\begin{figure}[!b]
    \centering
    \includegraphics[width=\columnwidth]{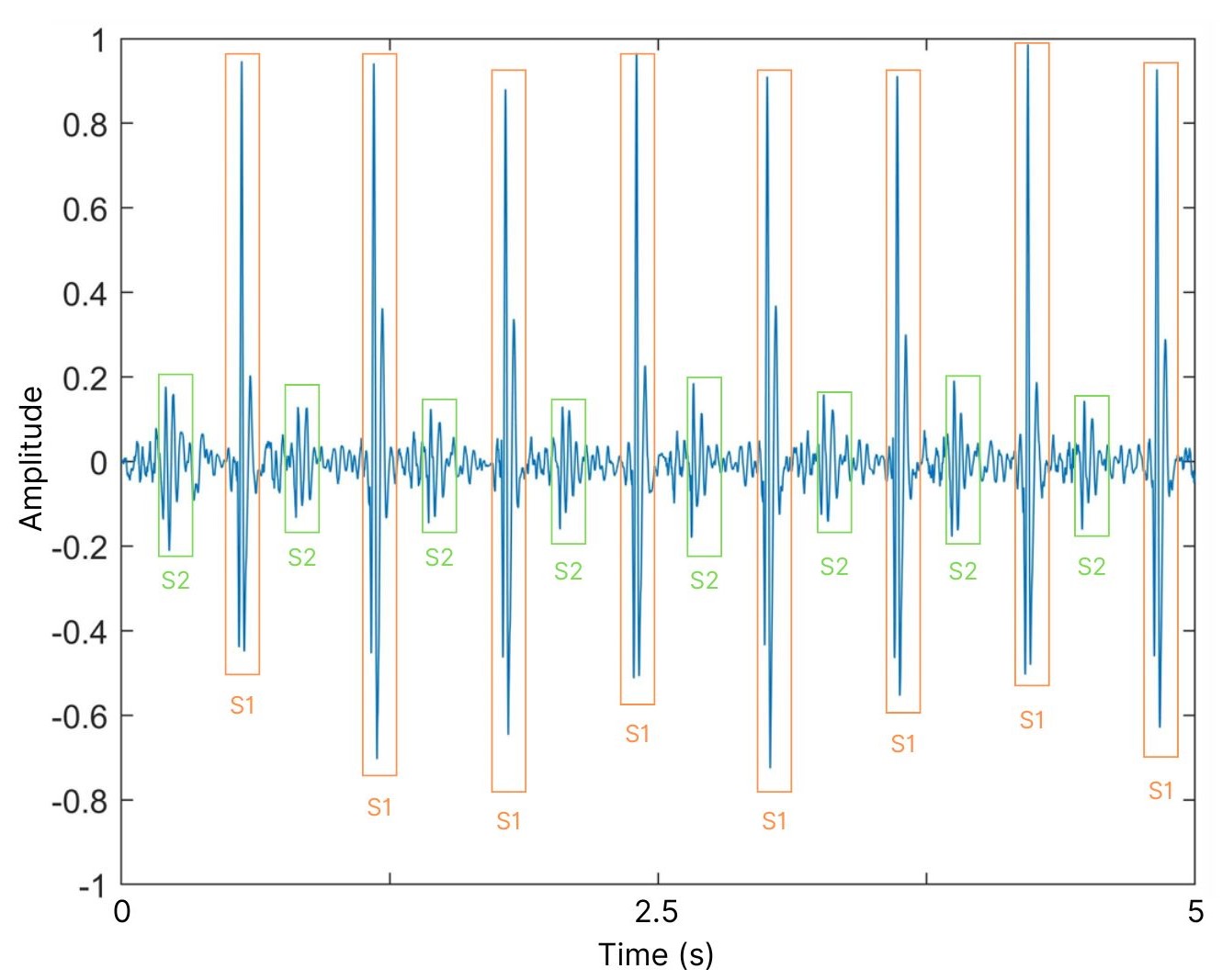}
    \caption{A Typical PCG Signal Segment (Diseased - MD (MR and AR)) from the BMD-HS Dataset in Bell Mode Filtering}
    \label{BMD-HS_data}
\end{figure}

\begin{figure*}[!ht]
    \centering
    \includegraphics[width=\textwidth]{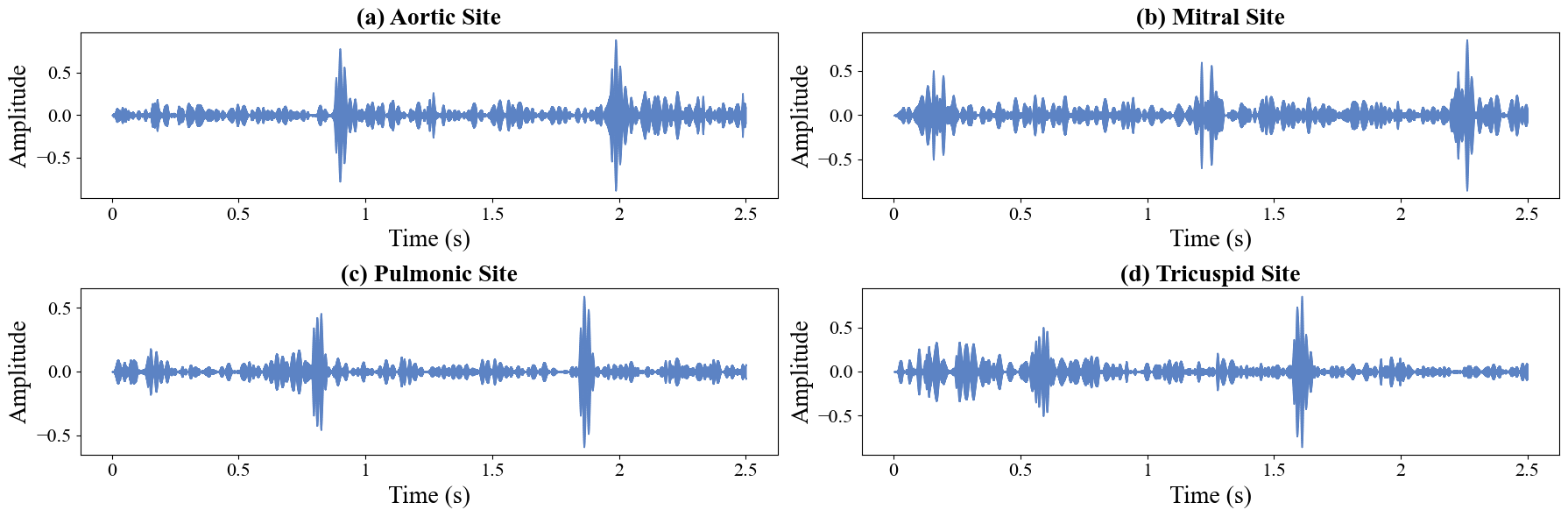}
    \caption{PCG Plot for 2.5 Seconds of the AR Class at Different Auscultation Sites: (a) Aortic Site, (b) Mitral Site, (c) Pulmonic Site, (d) Tricuspid Site}
    \label{aus}
\end{figure*}

\par 
Data from subjects admitted to the cardiac surgery ward (both male and female) of the National Institute of Cardiovascular Disease (NICVD), Dhaka, were collected for the diseased categories. Healthy subjects were represented by volunteers from the Department of Biomedical Engineering, Bangladesh University of Engineering (BUET). A clipped heart sound recording from the BMD-HS dataset is illustrated in Fig.~\ref{BMD-HS_data}.

\subsection{Study Procedure}
The cardiac condition of each subject was verified through an echocardiogram. The collection protocol utilized the diaphragm mode of a clinical-grade digital stethoscope, which was connected to a laptop via bluetooth. Data were recorded in both sitting and supine positions from four significant auscultation sites: mitral site, tricuspid site, pulmonic site, and aortic site. These sites are optimal for detecting the heart sounds related to the targeted diseases. Consequently, eight recordings of 20 seconds each were obtained from every subject. For male subjects, recordings were performed on bare skin, while for female subjects, a single layer of clothing was deemed acceptable. Characteristic plots from different auscultation sites for a patient with aortic regurgitation are shown in Fig.~\ref{aus}, displaying S1 and S2 heart sounds and murmurs, which serve as evidence of abnormalities. The rhythmic variations observed at these sites have been physical tools for diagnosing cardiac diseases \cite{pelech2004physiology}.
\par
\begin{figure}[ht!]
    \centering
    \includegraphics[width=\columnwidth]{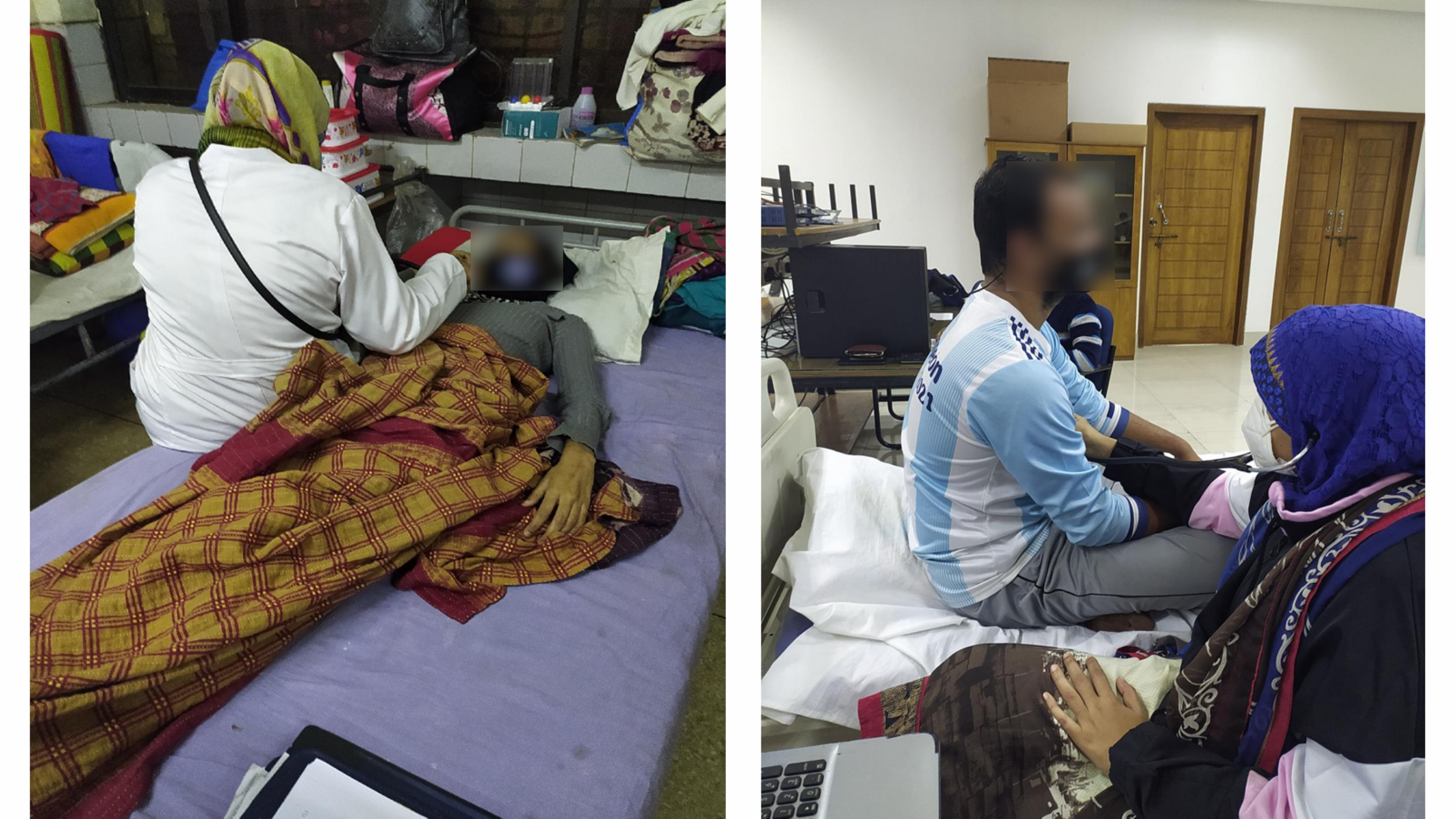}
    \caption{A Snapshot During Data Collection from Patients at NICVD (Left) and Healthy Volunteers in the Laboratory (Right)}
    \label{data_NICVD}
\end{figure}

 
PCGs were re-recorded if any ambiguity or noise was detected in the signal. Simultaneously, relevant metadata were collected from subjects using a dedicated app on a tablet, with all information kept confidential. A snapshot of the data collection process at NICVD and in the laboratory is presented in Fig.~\ref{data_NICVD}.
\par 
Due to COVID-19 conditions, PCGs of healthy control subjects were recorded in the laboratory under the same collection protocol. To ensure consistency with the hospital environment, a recorded background noise from NICVD was played during data collection for healthy subjects. This background noise had been previously recorded at NICVD to replicate the environmental conditions accurately. 

\subsection{Hardware and Software Utilized}
The primary hardware employed for data recording was the 3M™ Littmann® Model 3200 digital stethoscope~\cite{littmann}, which features a sampling frequency of 4 kHz. This stethoscope includes a wireless dongle that can be easily connected to a PC or laptop for use with its corresponding visualization software, 3M™ Littmann® StethAssist™~\cite{stethassist}. Within this software, raw recordings in .zsa' format can be saved as .wav' files, with options to adjust filter modes (Bell, Diaphragm, Extended).

Patient metadata were collected using the android app Open Data Kit Collect (ODK Collect) on a tablet.



\begin{figure*}[!t]
    \centering
    \includegraphics[width=\textwidth]{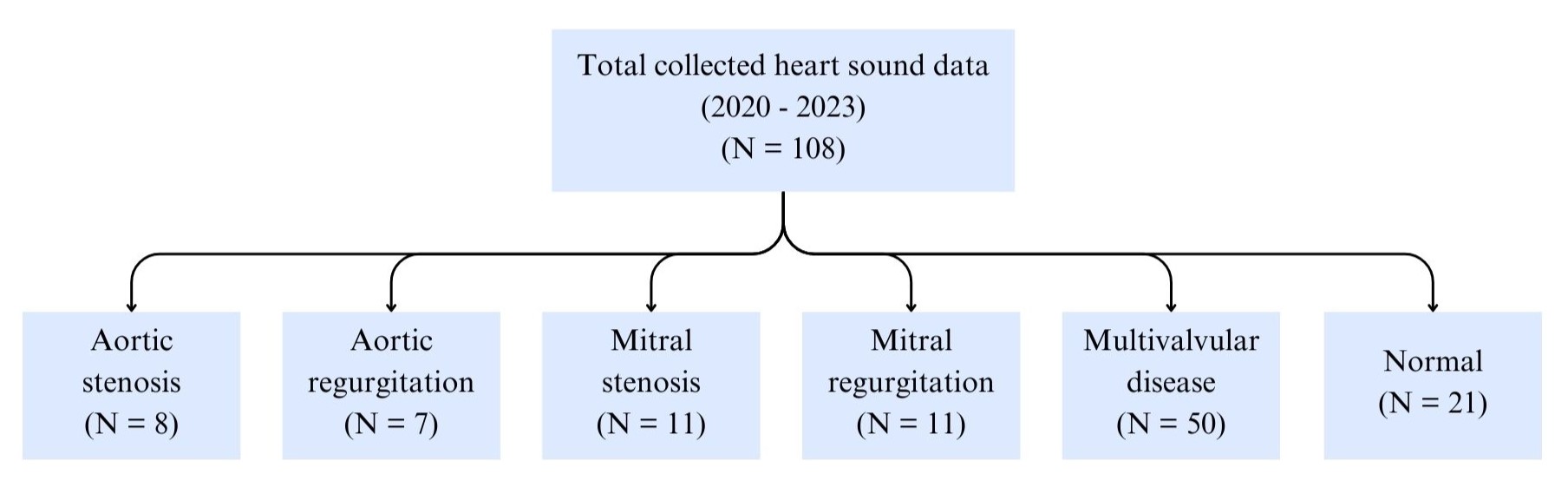}
    \caption{Current Demographics of the BMD-HS Dataset for Different Classes}
    \label{flow}
\end{figure*}

\subsection{Subject Inclusion Criteria}
Eligibility for data collection was determined by a clinical collaborator appointed for this study. Given that an echocardiogram is the gold standard for diagnosing valvular heart diseases (VHDs), patients were required to have confirmed diagnoses of the specific disorders targeted for the dataset through echocardiogram reports. Similarly, normal control subjects were required to verify the healthy condition of their cardiovascular system by submitting an echocardiogram report. All subjects were adults and affirmed that no false or misleading metadata were provided. The criteria for subject eligibility are as follows:

\begin{itemize}
    \item \textbf{Patient Criteria:} Patients from any age group were considered, with an emphasis on maintaining gender balance. Heart sound collection focused on patients diagnosed with one of the four targeted valvular diseases: aortic stenosis, aortic regurgitation, mitral stenosis, or mitral regurgitation.
    \item \textbf{Echocardiography:} The primary focus was on obtaining reliable data from patients with documented echocardiography reports. Adherence to this criterion was essential for verification, with confirmation provided by the medical collaborator.
    \item \textbf{Complete Information:} Participants who provided all necessary information for metadata were selected. This ensured the inclusion of comprehensive and accurate data in the study. Metadata primarily consisted of demographic information such as age and gender, as well as lifestyle details such as smoking habits and living area.
\end{itemize}

\subsection{Ethical Considerations}
This data collection study was approved by the NICVD Institutional Review Board (IRB) of NICVD\footnote{IRB Protocol No. NICVD/Academic/Study/2016-2017/6047}. Free and informed consent was obtained in written form from all subjects, and the study's nature, risks, and objectives were clearly communicated. All collected information, including heart sounds and patient metadata, is preserved with strict confidentiality. Access to this information is restricted to authorized research personnel, ensuring proper authentication.

\subsection{Data Annotation Strategy}
To ensure usability, the data annotations were designed to be precise, well-structured, and clear. A comprehensive annotation approach was adopted to allow detailed descriptions of heart sounds, associated conditions, and related problems. The following steps outline the annotation strategy:

\begin{itemize}
  \item A naming convention that includes the valvular disease type, patient ID, patient position during the recording, and the auscultation site was followed. This convention ensures that each file clearly identifies the patient, the specific illness, the patient’s position during the recording, and the site of the heart sound.
  \item The presence of a disease was indicated with binary values (1 for presence, 0 for absence). For instance, if aortic stenosis was present, it was marked as 1, with other diseases marked as 0. Normal patients were indicated with 1 for "normal" and 0 for others. In cases of multivalvular disease, multiple 1s were used. For example, a patient with severe mitral and aortic stenosis would have 1s for both MS and AS, with 0s for other diseases.
  \item Information on age, gender, smoking habits, and living area was included in the annotation to understand their impact on the disease and their interrelations.
\end{itemize}
Additionally, each annotation file was supplemented with an echocardiograph report to validate the data and confirm the presence of specific diseases.


\begin{table*}[!t]
\caption{DATA DEMOGRAPHICS IN TERMS OF GENDER, SMOKING STATUS AND INHABITANCE}
\label{demographics_table}
\centering
\renewcommand{\arraystretch}{1.5} 
\begin{tabular}{c c c c c c c c}
\hline
\multirow{2}{*}{\textbf{Category}} & \multirow{2}{*}{\textbf{Subcategory}} & \multicolumn{6}{c}{\textbf{Classes}} \\
& & \textbf{AS (N=8)} & \textbf{AR (N=7)} & \textbf{MR (N=11)} & \textbf{MS (N=11)} & \textbf{MD (N=50)} & \textbf{N (N=21)} \\ \hline
\multirow{2}{*}{\textbf{Gender}} & Male & 5 (62.5)\% & 5 (71.4\%) & 6 (54.5\%) & 45 (5.5\%) & 29 (58\%) & 12 (57.9\%) \\ 
& Female & 3 (37.5\%)& 2 (28.6\%) & 5 (45.5\%) & 6 (54.5\%)& 21 (42\%) & 9 (42.1\%)\\  \hline      
\multirow{2}{*}{\textbf{Smoking Status}} & Smoker & 3 (37.5\%) & 4 (57.1\%) & 3 (27.3\%) & 3 (27.3\%) & 19 (38\%) & 0 (0\%) \\
& Non-smoker & 5 (62.5\%) & 3 (42.9\%) & 8 (72.7\%) & 8 (72.7\%) & 31 (62\%) & 21 (100\%) \\ \hline
\multirow{2}{*}{\textbf{Living Area}} & Urban & 2 (26.6\%) & 3 (42.9\%) & 3 (27.3\%) & 4 (36.4\%) & 27 (54\%) & 21 (100\%) \\
& Rural & 6 (73.4\%) & 4 (57.1\%) & 8 (72.7\%) & 7 (63.6\%) & 23 (46\%) & 0 (0\%) \\ \hline
\end{tabular}
\end{table*}



\section{Demographic Analysis of Valvular Heart Disease}
Following all the criteria, heart sound data for 108 subjects was acquired at the end of our data collection journey, shown with a flow diagram in Fig.~\ref{flow}. Accordingly, the following demographics are now represented in the BMD-HS dataset: (a) AR (7); (b) AS (8); (c) MR (11); (d) MS (11); (e) MD (50); and (f) Normal (21). Among the 108 subjects, 46 are female, and 62 are male.

\subsection{Gender-Based Distribution}

From Table~\ref{demographics_table}, it can be observed that certain conditions, such as aortic stenosis and aortic regurgitation, exhibit a notable gender bias, with a higher percentage of male patients. Aortic stenosis, in particular, is more common among older men. Aortic regurgitation affects 71.4\% of males and 28.6\% of females. Similarly, aortic stenosis is diagnosed in 62.5\% of males, indicating a greater prevalence among men. In contrast, mitral valve disorders, including mitral stenosis and mitral regurgitation, show a generally balanced gender distribution, with no consistent gender disparity. For mitral regurgitation, 54.5\% of patients are men, and 45.5\% are women, suggesting a slight male predominance. Conversely, for mitral stenosis, 54.5\% of patients are women, and 45.5\% are men, indicating a slightly higher incidence among females.

\subsection{Age-Based Distribution}

The majority of patients are concentrated in the 20-25 and 35-45 age ranges. Notably, there is a higher frequency of patients aged around 22 and 40, which indicates a significant presence of both younger adults and middle-aged individuals within the dataset. Furthermore, a smaller but distinct number of patients are distributed across the extremes of the age range, particularly in the 13-20 and 60-70 age groups.


\subsection{Geographical Distribution}
The prevalence of valvular disease diagnoses varies between urban and rural areas due to factors such as lifestyle, socioeconomic status, access to healthcare, and the incidence of risk factors. Limited access to healthcare and specific lifestyle factors contribute to a higher occurrence of valvular disease in rural populations. As indicated in Table~\ref{demographics_table}, 57.1\% of aortic regurgitation cases are found in rural patients, compared to 42.9\% in urban patients. Aortic stenosis is diagnosed in 73.4\% of rural individuals and 26.6\% of urban individuals. In the case of mitral regurgitation, 72.7\% are rural patients, while 27.3\% are urban patients. Similarly, mitral stenosis is more prevalent in rural settings, with 63.6\% of cases compared to 36.4\% in urban areas.

\subsection{Impact of Smoking on Valvular Disease Incidence}
Table~\ref{demographics_table} presents the data distribution for smoker and non-smoker subjects. Among patients with aortic regurgitation, approximately 57.1\% were smokers, while the remaining patients did not smoke. In patients with aortic stenosis, 62\% were non-smokers, compared to approximately 37.5\% who were smokers. Smoking was present in about 27.3\% of patients with mitral regurgitation and 27.3\% of patients with mitral stenosis. Smokers constituted about 38\% of patients with multivalvular disease. It is important to note that these patients smoked before being diagnosed with valvular disease.

\subsection{Severity in Multivalvular Heart Disease}
Due to elements such as advancing calcification, structural anomalies, or the compounding effects of various valve diseases, aortic stenosis may be more common and severe in multivalvular disease. About 36\% of multivalvular diseased patients have severe aortic stenosis.


\begin{figure}[!h]
    \centering
    \includegraphics[width=\columnwidth]{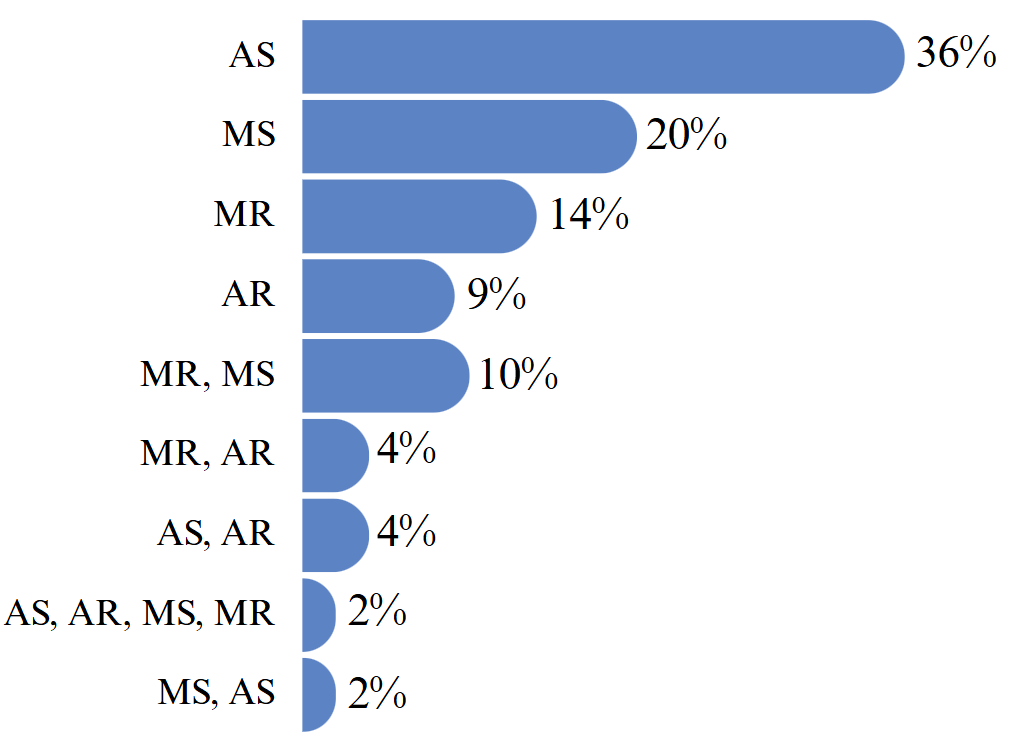}
    \caption{Severity Factors Found for Valvular Diseases}
    \label{multihd}
\end{figure}

As seen in Fig.~\ref{multihd}, mitral stenosis can be significant in multivalvular disease, its severity may vary based on the underlying etiology and anatomical factors. About 20\% of multivalvular diseased patients have severe mitral stenosis. The severity of mitral regurgitation in multivalvular disease can depend on factors such as the extent of valve damage, prolapse, or annular dilation. Among multivalvular heart disease patients about 14\% patients have severe mitral regurgitation. In multivalvular disease, the severity of aortic regurgitation may vary depending on factors such as valve morphology, annular dilation, or the presence of other valve abnormalities. About 8\% of patients have severe aortic regurgitation.

\begin{table*}[!t]
  \centering
  \begin{tabular}{c||c}
  \includegraphics[width=0.4\linewidth, height=5cm]{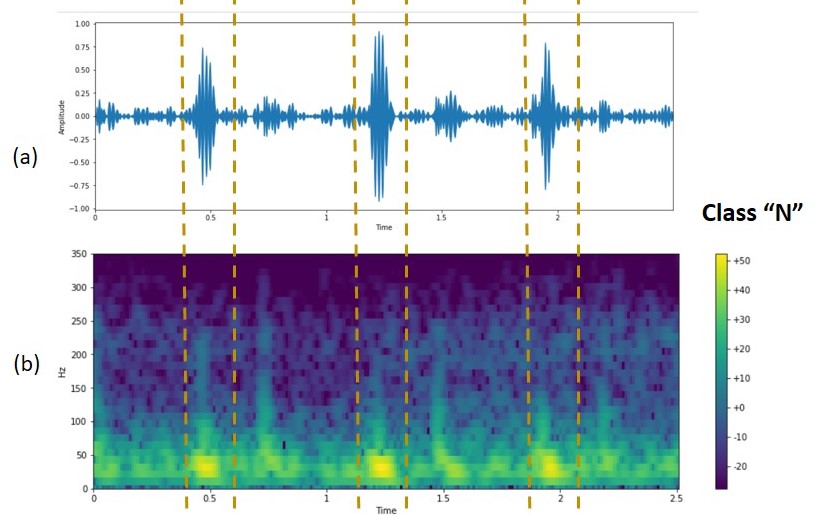} &
    \includegraphics[width=0.4\linewidth, height=5cm]{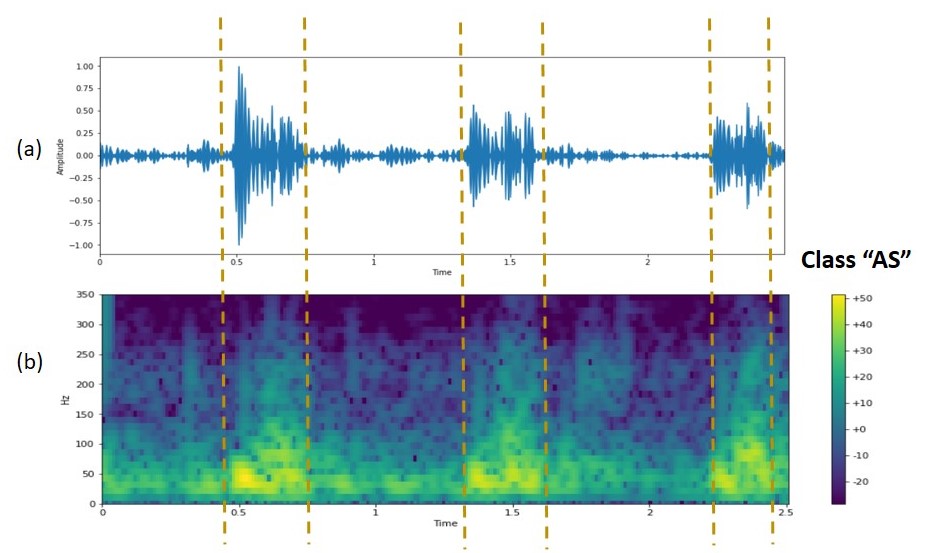} \\

    \textbf{Normal} & \textbf {Aortic Stenosis} \\
    \hline
    \includegraphics[width=0.4\linewidth, height=5cm]{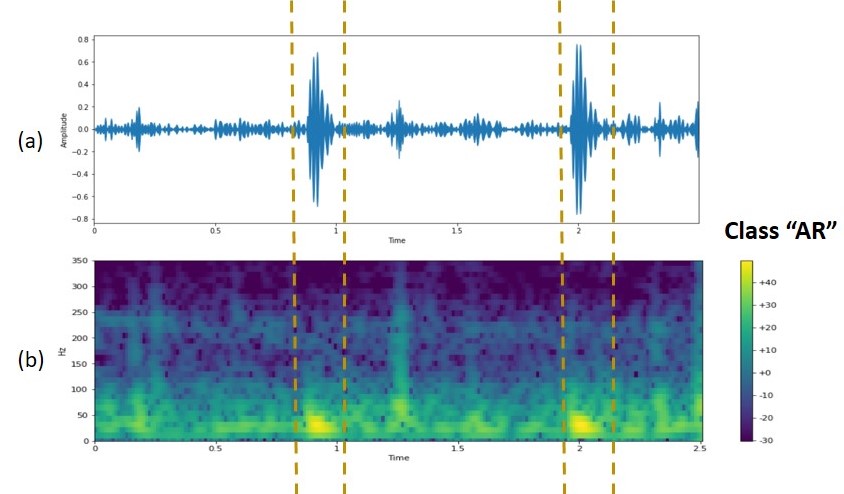} &

    \includegraphics[width=0.4\linewidth, height=5cm]{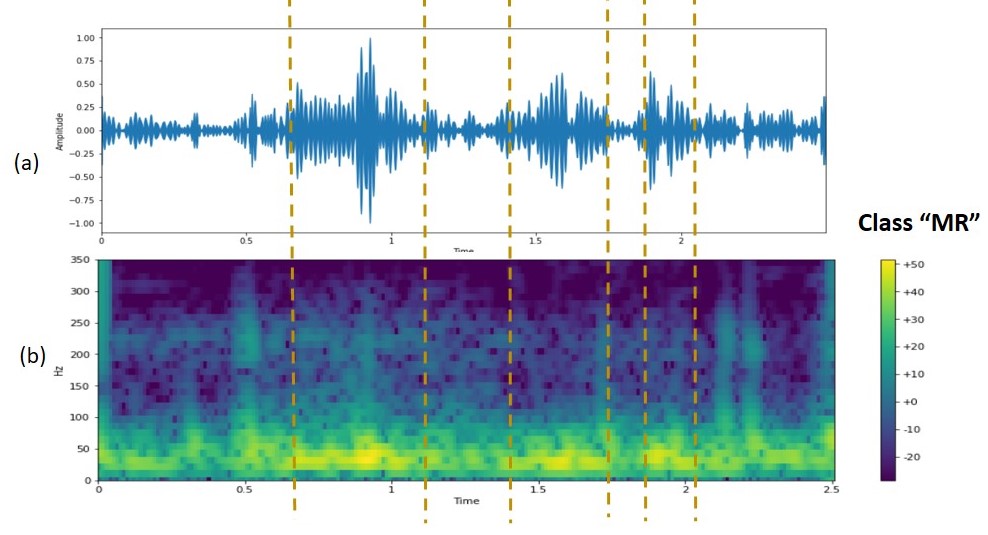} \\
    \textbf{Aortic Regurgitation} & \textbf{Mitral Regurgitation}\\
\hline
    \includegraphics[width=0.4\linewidth, height=5cm]{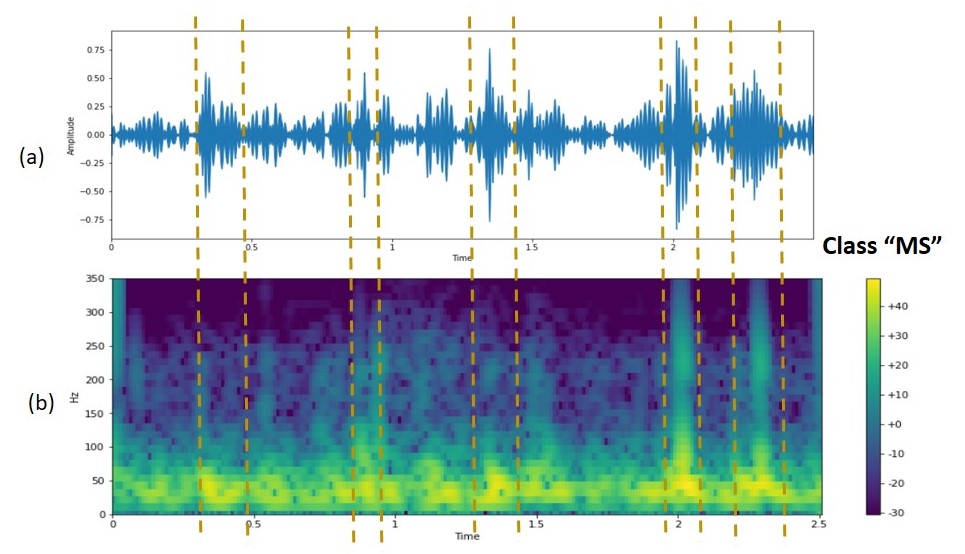} &
  
    \includegraphics[width=0.4\linewidth, height=5cm]{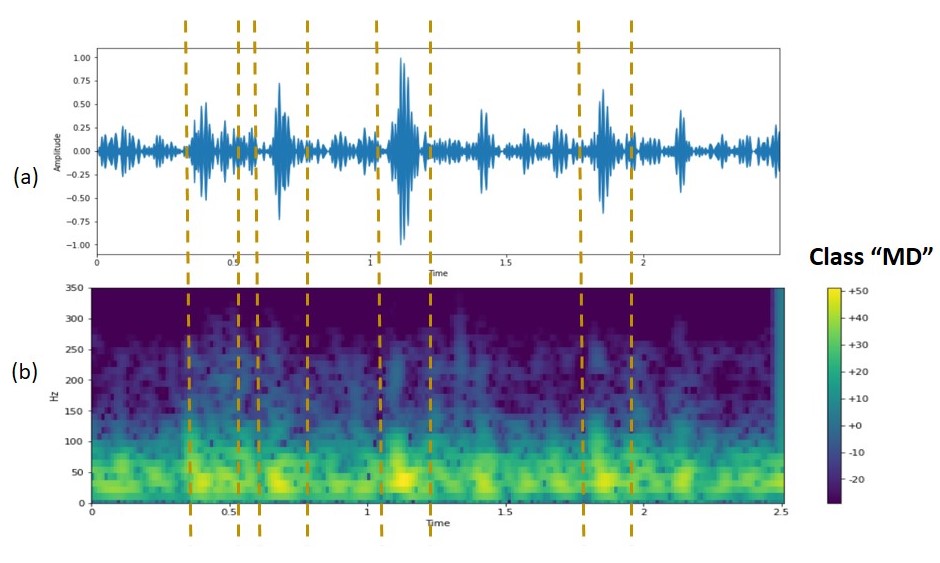} \\
    \textbf{Mitral Stenosis} & \textbf{Multi Disease}\\
    \hline
  \end{tabular}
  \caption{Compilation of Six Classes of Heart Sound Data for Temporal and Spectral Analysis}
  \label {table:spectral}
\end{table*}

\begin{figure}[!t]
    \centering
    \includegraphics[width=0.45\textwidth]{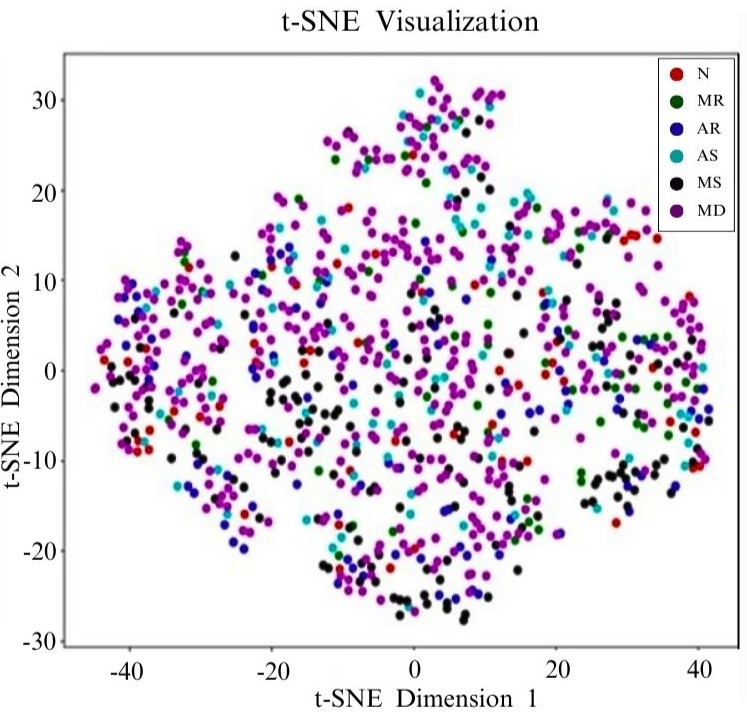}
    \caption{t-Stochastic Neighbour Embedding (t-SNE) Visualization of Normal (N) Class and Four CVDs Classes: Aortic Stenosis (AS), Aortic Regurgitation (AR), Mitral Regurgitation (MR), Mitral Stenosis (MS) Heart Sound Data}
    \label{fig:tsne}
\end{figure}

\section{Categorical Data Analysis}
The dataset employed in this study encompasses six distinct classes of phonocardiogram (PCG) signals, specifically Normal (N), Aortic Stenosis (AS), Aortic Regurgitation (AR), Mitral Stenosis (MS), Mitral Regurgitation (MR), and Multi Disease (MD) heart sound data. These classes were meticulously analyzed through both time-domain and frequency-domain methods to distinguish the characteristic features of each heart sound. Notably, periodic fluctuations associated with primary heart sounds, S1 and S2, were observed, with cardiac cycle events such as valve closures denoted by dashed vertical lines. A time-frequency representation of the PCG signal was also produced, where color intensity correlates with the magnitude (in dB) at each frequency and time point. Bright yellow regions signify higher magnitudes, thereby facilitating the analysis of heart sound frequency content over time.
The data analysis for all six classes, graphically summarized in Table~\ref{table:spectral}, is elaborated upon below:

\begin{itemize}
\item For the Normal (Class-N) category, a temporal and spectrographic analysis reveals that S1 and S2 frequencies are primarily between 20 and 100 Hz, with S1 having a notably longer duration than S2. The onset of diastolic and systolic phases is associated with these elevated frequencies.
\item In the Aortic Stenosis (Class-AS) category, the duration of S1 and S2 is extended, and a specific murmur, ejected post-S1, is identified in both time and frequency domains. This murmur extends to nearly 200 Hz, a characteristic feature of the crescendo-decrescendo pattern.
\item The Aortic Regurgitation (Class-AR) category is distinguished by a narrower frequency band compared to the Normal class, ranging from 20 Hz to approximately 40 Hz, attributed to an early diastolic decrescendo murmur that follows S2.
\item For the Mitral Regurgitation (Class-MR) category, the frequency range extends beyond that of a normal heart sound, including a pansystolic murmur up to 100 Hz. Significant spectral components, indicated by brighter colors, are concentrated below 150 Hz.
\item The Mitral Stenosis (Class-MS) category exhibits a murmur within the middle of the diastolic phase, lasting until just before the first cardiac sound (S1) and following an opening snap. The spectral analysis reveals a lower frequency spectrum ranging from approximately 30 Hz to 2000 Hz, indicative of reduced flow velocity and turbulence due to the stenotic tricuspid valve.
\item The Multi Disease (Class-MD) category presents a complex signal profile, incorporating features from multiple heart diseases. No consistent frequency range is observed for S1 and S2, with S1 spanning 20 to 150 Hz and S2 exceeding 200 Hz. The stronger presence of S2 and visible murmurs at various frequencies confirm the coexistence of multiple cardiovascular conditions.
\end{itemize}

\section{Data Distribution}

The dataset comprises 864 phonocardiogram (PCG) recordings, each of 20-second duration, collected from 87 individuals with valvular conditions and 21 individuals with normal heart conditions. The recordings represent four specific heart diseases and one multiclass category, resulting in six distinct classes. The dimensional reduction and cluster identification for this dataset were analyzed using a t-SNE plot, facilitating the understanding of data separability and aiding in the identification of decision boundaries crucial for classification tasks.

 Fig.~\ref{fig:tsne} illustrates the t-SNE plot, which shows that the classes do not form distinctly separate clusters. The feature space represented by t-SNE reveals significant overlap between classes, with data points dispersed across the plot. However, loose clustering is observed for classes AS (blue) and AR (red), indicating that these classes may have more defined characteristics in the high-dimensional space. The varying density across the plot suggests that data points in denser regions share greater similarity in the feature space.

\section{Challenges in Data Collection and Preparation}

The application of phonocardiograms in computer-aided diagnoses for cardiovascular diseases (CVDs) is hindered by several challenges, despite extensive research in the field. These challenges include the necessity for versatile and resilient solutions that account for diverse populations, stethoscope types, and environmental conditions. Publicly available heart sound datasets often face issues such as inconsistent data duration, low recording counts, and variations in data collection and processing, coupled with the lack of comprehensive patient metadata. Furthermore, patient cooperation was occasionally hindered by repeated medical student visits, necessitating efforts to persuade patients to contribute their heart sounds. Discrepancies in data, such as conflicting reports on multiple heart diseases, posed additional classification challenges, complicating the decision-making process between following echocardiography reports or basing classifications on heart murmur quality.

\begin{figure*}[!t]
    \centering
    \includegraphics[width=\textwidth]{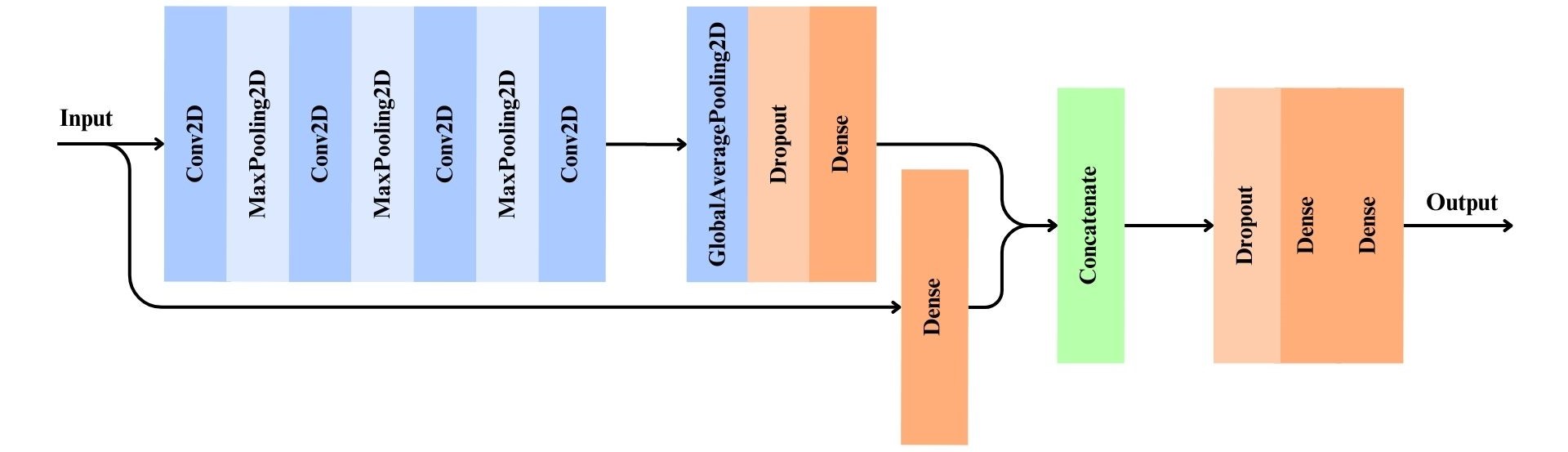}
    \caption{Architecture of the Primary Benchmarking Model}
    \label{model_architecture}
\end{figure*}

\section{Impact and Usability}

The collection of heart sound data from Bangladesh has profound social implications, particularly in the domains of healthcare, research, and technological development within developing countries. The dataset provides a comprehensive view of the current cardiovascular health state of the population across various demographics, offering a valuable resource for developing prescreening algorithms tailored to the specific needs of the region. The data collection process, spanning over 24 months, has been meticulously documented, and the dataset holds significant potential for advancing healthcare in Bangladesh, especially in rural areas where medical facilities are limited. This data can support remote diagnosis and monitoring efforts, while also contributing to the development of AI-based tools for early detection of cardiovascular diseases, a leading cause of mortality in the country. Additionally, this dataset contributes to the field of personalized medicine and informs public health policies by aiding in the allocation of resources and the formulation of strategies. Beyond its impact in Bangladesh, the dataset enhances the global understanding of cardiac disorders, addressing the current bias in medical datasets that are predominantly focused on Western populations. In conclusion, the collection of heart sound data from Bangladesh has far-reaching implications, elevating healthcare standards, advancing medical research, and broadening the global knowledge of cardiovascular health.

\section{Benchmarking and Evaluation}
\subsection{Model Architectures}
The primary model, illustrated in Figure \ref{model_architecture}, is designed to process two distinct inputs: Mel spectrograms derived from audio signals and metadata encoded as one-hot vectors representing posture and auscultation site information. The Mel spectrograms are passed through a series of four convolutional layers, each followed by max-pooling layers. These layers sequentially apply filters with ReLU activation, effectively reducing spatial dimensions while extracting essential features. A GlobalAveragePooling2D layer is then employed to condense the learned features by averaging each feature map across the entire spectrogram. The output is fed into a dense layer with 32 neurons, where ReLU activation is applied, and dropout is introduced to mitigate overfitting. Meanwhile, the metadata input is processed independently through a dense layer, also comprising 32 neurons. The outputs from both pathways are subsequently concatenated into a unified feature vector, which is further refined through an additional dense layer, followed by dropout. The final output layer, containing four neurons with sigmoid activation, is tailored for multi-label classification. The model is optimized using the Adam optimizer, with binary cross-entropy as the loss function and binary accuracy serving as the evaluation metric.

\begin{figure}[!h]
    \centering
    \includegraphics[width=0.4\textwidth]{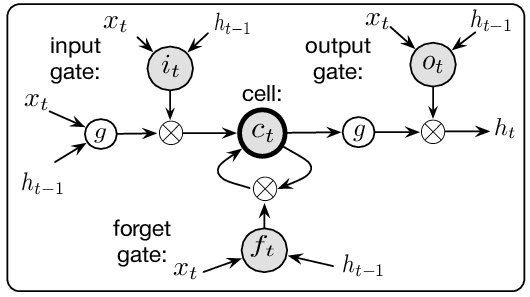}
    \caption{A General Block diagram of LSTM Architecture~\cite{Chen_2016}}
    \label{lstm}
\end{figure}

\begin{figure}[!h]
    \centering
    \includegraphics[width=0.4\textwidth]{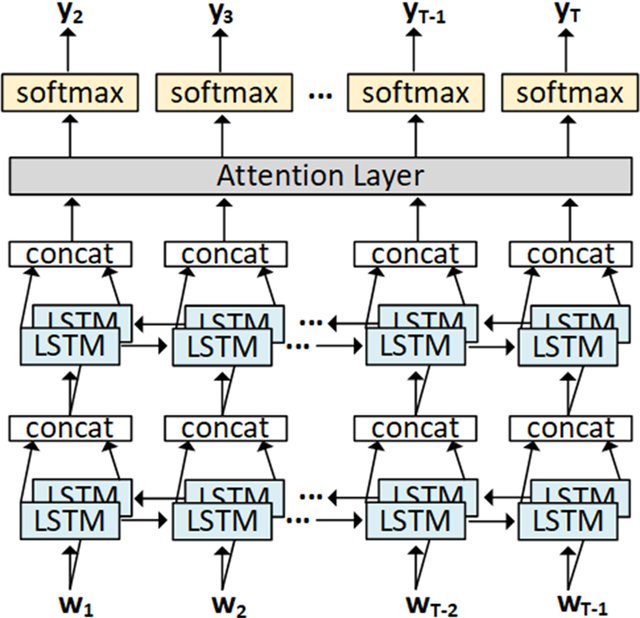}
    \caption{A General Block diagram of Bidirectional LSTM Architecture with Attention Layer~\cite{Wang2020}}
    \label{bilstm}
\end{figure}

\begin{figure}[!h]
    \centering
    \includegraphics[width=0.45\textwidth]{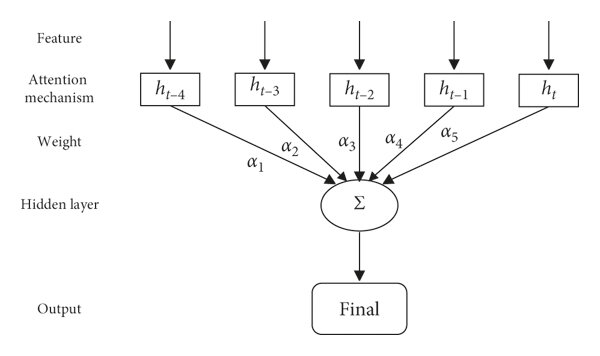}
    \caption{A General Block Diagram of GRU Architecture with Attention Layer~\cite{Xiong2021}}
    \label{gru}
\end{figure}

The LSTM (Long Short-Term Memory), Bidirectional LSTM, and GRU (Gated Recurrent Unit) models build upon this architecture by incorporating recurrent layers after the convolutional layers, as shown in Figures \ref{lstm}, \ref{bilstm}, and \ref{gru} respectively. These recurrent layers are instrumental in capturing the temporal dependencies within heart sound sequences.

The LSTM model, depicted in Figure \ref{lstm}, is a specialized recurrent neural network that excels at modeling temporal sequences and long-range dependencies. By leveraging memory cells capable of retaining information over extended periods, LSTM networks effectively address the vanishing gradient problem often encountered in standard RNNs. This architecture proves particularly effective in tasks where the sequence and duration of events play a critical role in prediction or classification, enabling the model to capture patterns that unfold over time.

In the Bidirectional LSTM model, illustrated in Figure \ref{bilstm}, the sequence data is processed through two stacked Bidirectional LSTM layers, each comprising 64 units, configured to output the entire sequence. This is followed by an attention mechanism, which assigns greater importance to significant features, allowing the model to focus on the most relevant aspects of the sequence. A global average pooling layer then reduces the sequence output to a single vector, which is combined with the processed metadata. As with the other models, the final output layer employs sigmoid activation to facilitate multi-class classification.

The GRU model, shown in Figure \ref{gru}, adopts a similar architecture, substituting GRU layers for LSTMs to efficiently capture temporal dependencies. As in the Bidirectional LSTM model, GRU layers are followed by an attention mechanism and global average pooling. The outputs from the GRU and metadata pathways are concatenated and passed through fully connected layers, with the final output layer utilizing sigmoid activation for multi-class classification.

\subsection{Training Methodology}

To train and evaluate the performance of various models on the BMD-HS dataset, the dataset was partitioned into training, validation, and testing subsets. Specifically, the dataset was divided into an 80:20 split, yielding 86 patients for training and validation, and 22 patients for testing. Each patient provided 8 recordings, resulting in a total of 688 recordings for the 86 patients utilized in the training and validation phases.

The Mel-spectrograms with 512 Mel bands was calculated for each audio file. The recordings were further split into training and validation subsets in an 80:20 ratio, yielding 550 recordings for training and 138 for validation. Given the limited number of patients, the model was trained using individual audio recordings as input, with the auscultation site and patient position provided as secondary inputs. Only the four disease labels were used during training; the N label was omitted, as it was inferred through the absence of any disease label.

The model was trained using the Adam optimizer with binary crossentropy loss. The specific hyperparameters for the Adam optimizer are detailed in Table~\ref{tab:hyperparameters}. Class weights were computed to address class imbalance, and early stopping and model checkpointing were utilized to prevent overfitting and to retain the best-performing model, respectively.

Given that the model was trained on individual audio recordings, an aggregation method was required to generate predictions at the patient level. For each patient, predictions were computed for all 8 recordings. The final prediction was determined by averaging the predictions across the 8 recordings and applying a threshold, optimized via brute-force to maximize the macro-averaged F1 score. This threshold was subsequently employed to generate predictions for the test set, enabling model evaluation.

\begin{table}[h]
    \centering
    \caption{Training Hyperparameters}
    \label{tab:hyperparameters}
    \begin{tabular}{cc}
        \hline
        \textbf{Hyperparameter} & \textbf{Value} \\
        \hline
        Learning Rate & 0.001 \\
        Beta 1 & 0.9 \\
        Beta 2 & 0.999 \\
        Epsilon & $1 \times 10^{-7}$ \\
        Batch Size & 64 \\
        Epochs & 500 \\
        Early Stopping Patience & 20 \\
        \hline
    \end{tabular}
\end{table}

\begin{table*}[!hb]
\centering
\caption{Comparison of Model Performance}
\renewcommand{\arraystretch}{1.5} 
\begin{tabular}{cccccc}
\hline
\textbf{Model} & \textbf{Accuracy} & \textbf{Sensitivity} & \textbf{Specificity} & \textbf{Macro-Averaged F1 Score} & \textbf{ICBHI Score} \\
\hline
Primary & 0.80 & 0.88 & 0.75 & 0.80 & 0.94 \\ 
Primary + BiLSTM & 0.70 & 0.76 & 0.69 & 0.69 & 0.83 \\ 
Primary + LSTM & 0.62 & 0.67 & 0.60 & 0.63 & 0.79 \\ 
Primary + GRU & 0.60 & 0.65 & 0.58 & 0.60 & 0.63 \\ \hline
\end{tabular}
\label{model_performance}
\end{table*}

\subsection{Evaluation Procedure}
To assess the performance of the models, various evaluation metrics were computed on the test set, including accuracy, sensitivity, specificity, and macro-averaged F1 score.

Accuracy, sensitivity, and specificity are essential metrics in binary classification that evaluate different aspects of a model's performance. To compute these metrics, the following quantities are first determined: True Positives (TP) represent the number of instances where both the actual and predicted labels are positive. True Negatives (TN) are cases where both the actual and predicted labels are negative. False Positives (FP) occur when the model incorrectly predicts a positive label for an instance that is actually negative, while False Negatives (FN) occur when the model incorrectly predicts a negative label for an instance that is actually positive.

Based on these quantities, the metrics are defined as follows:

\begin{equation} \text{Sensitivity} = \frac{\text{TP}}{\text{TP} + \text{FN}} \end{equation}

\begin{equation} \text{Specificity} = \frac{\text{TN}}{\text{TN} + \text{FP}} \end{equation}

\begin{equation} \text{Accuracy} = \frac{\text{TP} + \text{TN}}{\text{TP} + \text{TN} + \text{FP} + \text{FN}} \end{equation}

Sensitivity (also referred to as recall) measures the proportion of actual positive cases that are correctly identified by the model, reflecting its ability to detect positive instances. Specificity, on the other hand, measures the proportion of actual negative cases that are correctly identified, indicating how well the model avoids false positives. Accuracy provides an overall measure of the model's performance by calculating the proportion of all correct predictions (both positive and negative) out of the total number of predictions made.

The macro-averaged F1 score is a commonly used metric for multi-class classification tasks, especially when the class distribution is imbalanced. The F1 score is the harmonic mean of precision and recall, defined for each class $i$ as:

\begin{equation}
\text{F1}_{i} = \frac{2 \times \text{Precision}_{i} \times \text{Recall}_{i}}{\text{Precision}_{i} + \text{Recall}_{i}}
\end{equation}

where
\begin{equation}
\text{Precision}_{i} = \frac{\text{TP}_{i}}{\text{TP}_{i} + \text{FP}_{i}}
\end{equation}
\begin{equation}
\text{Recall}_{i} = \frac{\text{TP}_{i}}{\text{TP}_{i} + \text{FN}_{i}}
\end{equation}

The macro-averaged F1 score is calculated by averaging the F1 scores across all classes:

\begin{equation}
\text{Macro-Averaged F1 Score} = \frac{1}{N} \sum_{i=1}^{N} \text{F1}_{i}
\end{equation}

This metric gives equal weight to each class, making it suitable for evaluating the model's performance across all classes, regardless of their frequency.

In addition to these metrics, a modified ICBHI (International Conference on Biomedical and Health Informatics) score was calculated to provide a comprehensive evaluation of the model's performance on the specific task of classifying cardiac conditions for each patient. The modified ICBHI score is given by:

\begin{equation}
\text{ICBHI Score} = \frac{\text{Sensitivity}^\prime + \text{Specificity}^\prime}{2}
\end{equation}

where,
\begin{equation}
\text{Specificity}^\prime = \frac{C_{\text{AS}} + C_{\text{AR}} + C_{\text{MS}} + C_{\text{MR}}}{N_{\text{AS}} + N_{\text{AR}} + N_{\text{MS}} + N_{\text{MR}}}
\end{equation}

\begin{equation}
\text{Sensitivity}^\prime = \frac{C_{\text{N}}}{N_{\text{N}}}
\end{equation}

Here, $C_{\text{AS}}$, $C_{\text{AR}}$, $C_{\text{MS}}$, and $C_{\text{MR}}$ represent the correctly labeled cases of aortic stenosis, aortic regurgitation, mitral stenosis, and mitral regurgitation, respectively. Similarly, $N_{\text{AS}}$, $N_{\text{AR}}$, $N_{\text{MS}}$, and $N_{\text{MR}}$ denote the total number of cases for these conditions. $C_{\text{N}}$ and $N_{\text{N}}$ represent the correctly labeled and total cases of normal cardiac conditions, respectively.

\begin{figure}[!t]
    \centering
    \includegraphics[width=0.45\textwidth]{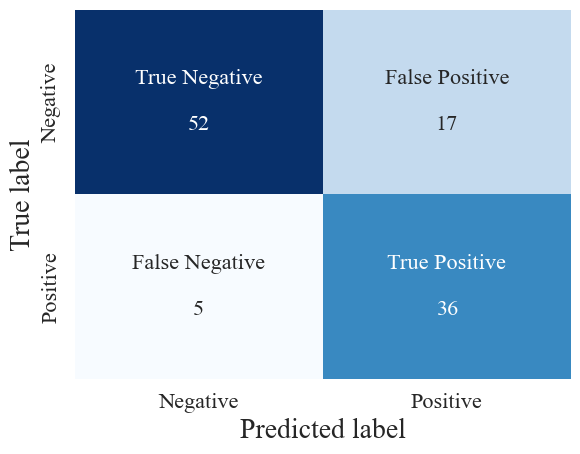}
    \caption{Confusion Matrix of the Primary Model’s Predictions on the Test Set}
    \label{fig:confusion}
\end{figure}

\subsection{Results}
The results from the performance comparison (Table \ref{model_performance}) indicate that the primary model outperformed the more complex recurrent models across all key metrics. This suggests that the simpler architecture, which focuses on extracting features from individual Mel-spectrograms and incorporates metadata through dense layers, is more effective at handling the classification task than the recurrent models. The confusion matrix for the primary model’s predictions is shown in Fig. \ref{fig:confusion}.

In contrast, the Bidirectional LSTM model, which introduces recurrent layers to capture temporal dependencies, showed a drop in performance. While it provided a reasonable sensitivity of 0.76, its lower specificity (0.69) and macro-averaged F1 score (0.69) indicate that the model struggled to maintain a balanced classification performance across all classes. This may suggest that the temporal features captured by the recurrent layers did not contribute positively to the overall classification task, possibly due to the limited amount of temporal variation in the input sequences.

The standard LSTM model performed even worse, with an accuracy of 0.62 and a macro-averaged F1 score of 0.63. The drop in specificity (0.60) suggests that this model was prone to false positives, which could have reduced its overall effectiveness. Similarly, the GRU model, which is designed to efficiently capture dependencies in sequential data, also underperformed with the lowest accuracy (0.60) and ICBHI score (0.63) among all the evaluated models. 

The poorer performance of the LSTM, Bidirectional LSTM, and GRU models, compared to the primary model, implies that the recurrent layers may not be necessary or optimal for this specific task. The primary model’s success, despite its simpler structure, suggests that the critical features for classifying the heart sound recordings are more effectively captured through convolutional operations and metadata integration rather than through the sequential modeling of the audio data. This finding aligns with the observation that heart sound recordings, particularly in this dataset, might not require extensive modeling of temporal dependencies, as the key discriminative features may be more localized and spatial in nature.

\section{Conclusions}
In this study, we introduced the BUET Multi-disease Heart Sound (BMD-HS) dataset, a rigorously curated collection of heart sound recordings aimed at enhancing the accuracy and reliability of cardiovascular disease (CVD) diagnosis through machine learning. The dataset addresses critical gaps in existing resources by offering multi-label annotations, standardized data collection methods, and echocardiogram-confirmed diagnoses. This dataset is particularly valuable for developing and benchmarking AI models tailored to complex heart sound classification tasks. Through comprehensive demographic analysis and benchmarking against various machine learning models, we demonstrated the dataset's potential in advancing automated CVD detection. The BMD-HS dataset not only provides a robust foundation for future research but also holds significant promise for improving diagnostic practices and healthcare outcomes, especially in resource-constrained settings.


\bibliographystyle{IEEEtran}

\bibliography{ref}
\end{document}